\RequirePackage{lineno}
\documentclass[aps,onecolumn,superscriptaddress,prl,floatfix,british,linenumbers]{revtex4}

\usepackage{amssymb}
\usepackage{amsmath}
\usepackage{babel}
\usepackage[sort&compress]{natbib}
\usepackage{topcapt}
\usepackage{pdfpages}

\usepackage{graphicx}
\usepackage{subfigure}

\usepackage[colorlinks,linkcolor=dred,urlcolor=dblue,citecolor=dgreen]{hyperref}
\usepackage{stmaryrd}
\usepackage{amsfonts}
\usepackage{mathrsfs}
\usepackage{txfonts}
\graphicspath{{figs/}}

\usepackage{xcolor}
\usepackage{tikz}
\usepackage{circuitikz}
\usetikzlibrary{backgrounds,fit,decorations.pathreplacing,calc,backgrounds}  

\usepackage{color}
\usepackage{pdfpages}
\definecolor{dred}{rgb}{.8,0.2,.2}
\definecolor{ddred}{rgb}{.8,0.5,.5}
\definecolor{dblue}{rgb}{.2,0.2,.8}
\definecolor{dgreen}{rgb}{.2,0.5,.2}

\usepackage{multirow}

\newcommand{\ket}[1]{| #1 \rangle}

\newcommand{\Tr}{\mathrm{Tr}}

\newcommand{\vecx}{\mathbf{x}}
\newcommand{\vecX}{\mathbf{X}}

\newcommand{\matA}{\mathbf{A}}
\newcommand{\opA}{\mathbf{A}}

\newcommand{\opD}{\mathbf{D}}

\def\be{\begin{eqnarray}}
\def\ee{\end{eqnarray}}

\usepackage{times}

\definecolor{Pr}{rgb}{0.4,0.3,0.9}

\begin{document}
\title{Optimizing a Polynomial Function on a Quantum Processor}

\author{Keren Li}
\thanks{These authors contributed equally to this work.}
\affiliation{State Key Laboratory of Low-Dimensional Quantum Physics and Department of Physics, Tsinghua University, Beijing 100084, China}
\affiliation{Shenzhen JL Computational Science and Applied Research Institute, Shenzhen 518109, China}

\author{Shijie Wei}
\thanks{These authors contributed equally to this work.}
\affiliation{Beijing Academy of Quantum Information Sciences, Beijing 100193, China}
\affiliation{State Key Laboratory of Low-Dimensional Quantum Physics and Department of Physics, Tsinghua University, Beijing 100084, China}

\author{Pan Gao}
\affiliation{State Key Laboratory of Low-Dimensional Quantum Physics and Department of Physics, Tsinghua University, Beijing 100084, China}

\author{Feihao Zhang}
\affiliation{State Key Laboratory of Low-Dimensional Quantum Physics and Department of Physics, Tsinghua University, Beijing 100084, China}

\author{Zengrong Zhou}
\affiliation{State Key Laboratory of Low-Dimensional Quantum Physics and Department of Physics, Tsinghua University, Beijing 100084, China}

\author{Tao Xin}
\affiliation{Shenzhen Institute for Quantum Science and Engineering,
Southern University of Science and Technology, Shenzhen 518055, China}

\author{Xiaoting Wang}
\email{xiaoting@uestc.edu.cn}
\affiliation{Institute of Fundamental and Frontier Sciences, University of Electronic Science and Technology of China, Chengdu, 610054, China}

\author{Patrick Rebentrost}
\email{pr@patrickre.com}
\affiliation{Centre for Quantum Technologies, National University of Singapore, Singapore 117543}

\author{Guilu Long}
\email{gllong@tsinghua.edu.cn}
\affiliation{State Key Laboratory of Low-Dimensional Quantum Physics and Department of Physics, Tsinghua University, Beijing 100084, China}
\affiliation{ Beijing National Research Center for Information Science and Technology and School of Information Tsinghua University, Beijing 100084, China}
\affiliation{Beijing Academy of Quantum Information Sciences,  Beijing 100193, China}
\affiliation{Frontier Science Center for Quantum Information, Beijing 100084, China}

\begin{abstract}
  The gradient descent method is central to numerical optimization and is the key ingredient in many machine learning algorithms. It promises to find a local minimum of a function by iteratively moving along the direction of the steepest descent. Since for high-dimensional problems the required computational resources can be prohibitive, it is desirable to investigate quantum versions of the gradient descent, such as the recently proposed [Rebentrost et.al.,2019]. Here, we develop this protocol and implement it on a quantum processor with limited resources. A prototypical experiment is shown with a 4-qubit Nuclear Magnetic Resonance quantum processor, which demonstrates the iterative optimization process. Experimentally, the final point converged to the local minimum with a fidelity above 94\%, quantified via full-state tomography. Moreover, our method can be employed to a multidimensional scaling problem, showing the potential to outperform its classical counterparts. Considering the ongoing efforts in quantum information and data science, our work may provide a faster approach to solving high-dimensional optimization problems and a subroutine for future practical quantum computers.
  \end{abstract}
\maketitle

\section{introduction}
A basic situation in optimization is the minimization or maximization of a polynomial subject to some constraints. As polynomials are in general non-convex and optimization is $NP$-hard, these problems cannot be solved accurately with efficient resource consumption\cite{2010-HomoPoly-Chesi}. As a special case for approximation algorithms,  homogeneous polynomial optimization has wide applications, for examples, signal processing, magnetic resonance imaging\cite{2008-PolyMRI-Ghosh}, data training\cite{2016-Polytraining-Blondel}, approximation theory\cite{2002-HomoPolyAproTheo-Kofidis} and material science\cite{2008-HomoPolyMat-So}. These scientific and technological problems are especially demanding in the present-day era of big data\cite{2010-HomoPoly-He}. The gradient algorithm, serving as one of the most fundamental solutions to non-convex optimization problems, lies at the heart of many machine learning methods, such as regression, support vector machines, and deep neural networks\cite{2019-MLhandbook-Geron,2019-Zhang-GforDP,2018-Manogaran-GLR,2018-wang-GSVD,2018-Du-GNN}. However, when dealing with large data sets, the gradient algorithm consumes tremendous resources and often pushes the current computational resources to their limits. 

Quantum computing promises ultra-fast computational capabilities by information processing via the laws of quantum mechanics\cite{2002-nielson-QCQI,1994-shor-shor,1996-Grover-Grover}. With the intrinsic advantages in executing certain matrix multiplication operations, quantum algorithms are proposed to enhance data analysis techniques under some circumstances. For example, phase estimation, quantum principal component analysis and the solver for linear system of equations can provide quantum advantages if the state preparation and readout procedure  can be efficiently realized\cite{2002-nielson-QCQI,2009-harrow-PRL-HHL}. As for the optimization with gradients, which is the central issue in this article, several works focusing on developing quantum versions\cite{2005-jordan-Qgradient,2019-Nathan-qGradient,2017-Hybrid-Li,2019-qGradient-schuld,2019-Rebentrost-Qgradient,2020-qGradient-Kerenidis} have been done.

Optimization, i.e., maximization or minimization, of a cost function can be attempted by the prototypical gradient algorithm iteratively. Let the cost function be a map  $f: \mathbb{R^N} \rightarrow \mathbb{R}$. Set an initial guess $\vecx^{(0)} \in \mathbb{R}^N$, then move it along the direction of the gradient

\begin{equation}
\vecx^{(t+1)} = \vecx^{(t)} \pm \eta\; \nabla f(\vecx^{(t)}),
\label{Eq:grad_algo}
\end{equation}
where $\vecx = (x_1, \dots, x_N)^T  \in \mathbb{R}^N$ and  $\eta $ is the learning rate. 
As the first experimental endeavor in this field and for the wide academic and industrial applications, in this work, an order-$2p$ homogeneous polynomial optimization with the spherical constraints $\sum_i ||x_i^2||=1$ is investigated, whose cost function is expressed as 

\begin{equation}
f(\vecx) = \sum\limits_{i_1,\dots,i_{2p}=1}^{N}    a_{i_1\dots i_{2p}}  x_{i_1}  \dots  x_{i_{2p}}.
\label{Eq:of1}
\end{equation}
The coefficients $a_{i_1\dots i_{2p}} \in \mathbb{R}$  can be reshaped to a $N^p\times N^p$ matrix $\matA$. And $f(\vecx)$ can be rewritten as $\frac{1}{2} \vecx^T\otimes ...\otimes  \vecx^T \matA   \vecx \otimes ...\otimes  \vecx$.
Simultaneously, $\matA$ is the linear summation  of tensor product of $N \times N$ unitary matrix $\matA_i^{\alpha}$, which represents as $\sum_{\alpha=1}^K \matA_{1}^{\alpha}\otimes...\otimes \matA_{p}^{\alpha}$. $K$ is the number of decomposition terms required to specify $\matA$ and $p$ is half the order of the cost function.
Therefore, in light of the previous work,
the gradient at $\vecx$ can be mapped into a matrix summation\cite{2019-Rebentrost-Qgradient}

\begin{equation}
\nabla f(\vecx) =  \sum_{\alpha=1}^K \sum_{j=1}^p   \big(\prod^p_{i=1 \atop i\neq j}\vecx^{T} \matA^{\alpha}_i \vecx \big)  \matA^{\alpha}_j \; \vecx ,
\label{Eq:deriv}
\end{equation}
With the amplitude encoding method, which encodes $\vecx$ as $|\vecx\rangle=\sum_{i=0}^{N-1} x_i|i\rangle$, the iterative equation Eq.~(\ref{Eq:grad_algo}) is interpreted as an evolution of $|\vecx\rangle$ with an operation of $\opD$,
\begin{eqnarray}\label{DD1}
|\vecx^{(t+1)}\rangle&=& |\vecx^{(t)}\rangle \pm \opD |\vecx^{(t)}\rangle,\nonumber\\
\opD&=&\sum_{\alpha=1}^{K} \sum_{j=1}^{p} (\prod_{i=1, i\neq j}^p \langle\vecx^{(t)}| \matA^{\alpha}_i|\vecx^{(t)}\rangle) \opA_{j}^{\alpha}
\end{eqnarray}
where $\opD$ is a parameter-dependent gradient operator. which in general is non-unitary.
Note that two methods to decompose $\matA$ are shown in supplementary note \textrm{I}(D)\cite{supple} and the subscripts $^{(t)}$ will be omitted in the remainder.

In this work, we develop the gradient algorithm and propose an experimental protocol to perform the gradient descent iterations, with a prototypical experiment to demonstrate the process to optimize polynomials on a quantum simulator. The gradient algorithm in [Rebentrost et.al.,2019] \cite{2019-Rebentrost-Qgradient} involves  phase estimation which  requires substantial circuit depth for currently available circuits, giving logarithmic error and polynomial gate scaling. Hence, the algorithm is difficult to implement with current techniques on a quantum platform. Instead of phase estimation, our method uses the linear combination of unitaries to realize the gradient descent iterations. It provides a gate-based circuit only comprising of standard quantum gates, which is experimental friendly and implementable in current quantum techniques. Our protocol needs two copies of a quantum state $ |\vecx\rangle $ to produce the next quantum state  at each iterative step, instead of  multiple copies which is linearly depending on the order of objective function in the  previous algorithm \cite{2019-Rebentrost-Qgradient}. 
The product of decomposition terms $K$ and the order $p$ is an important indicator to determine the efficiency of our protocol. The protocol can be especially beneficial in cases  when there is an explicit decomposition of $A$ with comparably small $Kp$ and $\matA_i^{\alpha}$ is Pauli product matrix, calculating the gradient with the $\mathcal{O}(Kp\times \log(N))$ depth circuits. Moreover, the experiment benefits from the protocol as only two copies are required for optimization of each iteration. Therefore, given the unrivaled degree of nuclear magnetic resonance (NMR) quantum control techniques\cite{2005-ReviewNMR-Vandersypen,2010-reviewQC-Ladd}, this homogeneous polynomial optimization is conducted with a 4-qubit system encoded in a molecule of crotonic acid and a quantum state in the vicinity of the local minimum is iteratively obtained with high fidelity. Finally, Multi-Dimensional Scaling (MDS) problems are introduced as a potential application of this protocol.

\section{results}
\subsection*{Experimental protocol}
For the convenience to implement the evolution of the gradient operation, $\opD$ (shown in Eq.(\ref{DD1})) is rewritten as

\begin{eqnarray}
\opD=\sum_{\alpha=1}^{K} \sum_{j=1}^p M^{\alpha} \dfrac{\opA_{j}^{\alpha}}{b_{j}^{\alpha}} = \sum_{m=0}^{Kp-1}   c_{m}\matA_{m} 
\end{eqnarray}
where $\matA_{m}$ and $\matA^{\alpha}_j$ are the same if $m=p(\alpha-1)+ j-1$. In addition, $b_{j}^{\alpha} =\langle\vecx| \matA^{\alpha}_j|\vecx\rangle$ and $M^{\alpha}=\prod_{i=1}^p b_{i}^{\alpha}$.

\begin{figure}[!ht]
  \centering
  \includegraphics[width= 0.95 \columnwidth]{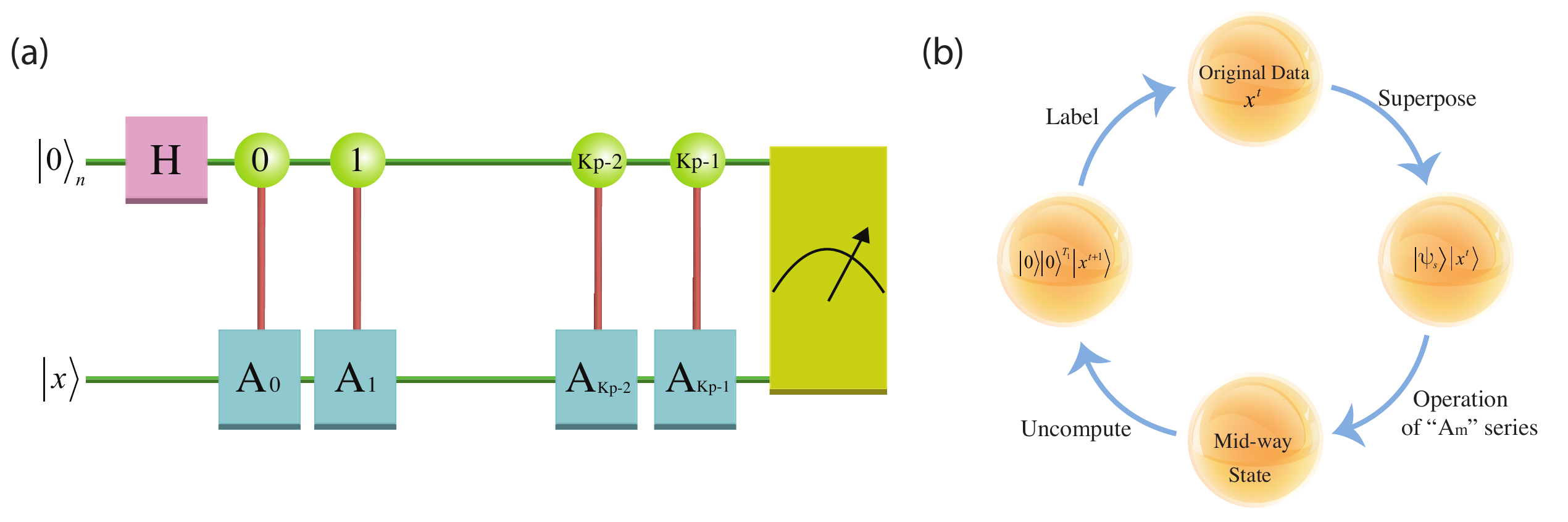}
  \caption{Circuits for implementing the quantum gradient algorithm. $|\vecx\rangle$ denotes the input state of the work system, and  ancillary systems are $ T_{1}+1 $ qubits in the  $|0\rangle|0\rangle^{\otimes T_{1}}$ state, where $ T_{1}=\log_{2}(Kp) $.  The squares represent unitary operations and the circles represent the state of the controlling qubit. (a) Parameter circuit. $b_{j}^{\alpha}$ can be obtained with $\langle\sigma_x\rangle$ on register $s$ when register $d$ is on $|m=p(\alpha-1)+ j\rangle $. (b) Iteration circuit includes three steps: initialization, $\opD$ application, combination.}
  \label{fig1}
  \end{figure}

As shown in Fig.\ref{fig1}, to implement the quantum gradient algorithm , two specified circuits are involved. Parameters such as $c_m (m=0...Kp-1)$ can be obtained by the parameter circuits in Fig.\ref{fig1}(a). This circuit evolves the system

\begin{eqnarray}
|0\rangle_s|0\rangle_d^{T_{1}}|\vecx\rangle  \rightarrow \frac{1} {\sqrt{2}} \left ( |0\rangle_s|0\rangle_d^{T_{1}}|\vecx\rangle +\sum_{m=0}^{2^{T_1}-1} \frac{1} {\sqrt{2^{T_1}}}|1\rangle_s|m\rangle_d A_m|\vecx\rangle\right),  
\end{eqnarray}
where $T_1$ is the integer that satisfies $2^{T_1}= Kp$. When the ancillary system $d$ is in state $ |m\rangle $, where $m=p(\alpha-1)+ j-1$, $ b_{j}^{\alpha}$ can be obtained on the ancillary system $s$ with $\sigma_x$ basis. Thus $ M^{\alpha}$ ($\alpha$=1...$K$),  as well as $c_m$ can be calculated once the $m$ traverses $[0, Kp-1]$.

The iteration circuit, which is shown in Fig.\ref{fig1}(b), is to generate the iterative state with $\opD$. 
The ancillary system $s$ is the first ancillary system which is for the linear combination of two terms in Eq.(\ref{DD1}) and $d$ is second one which is for the implementing $\opD$ with linear combination of operators. 
In our protocol, the first thing is to involve the minus signs in $c_m$ in the unitary operator $\matA_m$ to make  $c_m$ positive. After dealing with the minus signs in Eq.(\ref{DD1}),  the iterative equation is rewritten as 

\begin{equation}\label{Exp_state}
|\vecx'\rangle= \left (|\vecx\rangle + \sum_{m=0}^{Kp-1} c_m \matA_m|\vecx\rangle \right).
\end{equation}
where $|\vecx'\rangle$ is the state for the next step. The entire iteration circuit to $|\vecx'\rangle$ can be implemented via following four steps: 

Step 1: For the work register, the amplitude encoding state  $|\vecx\rangle$ should be efficiently prepared. In general, for the first iteration, some easy-access states can be our initial states, such as tensor product states. For the following iterations, the output of the last iteration can generate what we want. Hence in this situation, time complexity can be ignored. In the case of preparing a particular $\log(N)$-qubit input $|\vecx\rangle=\sum_k c_k|k\rangle$, we employ the amplitude encoding method in Ref \cite{grover2002creating,soklakov2006efficient}. It shows that if $c_k$ and $P_k=\sum_k |c_k|^2 $ can be efficiently calculated by a classical algorithm, constructing this particular state takes $O[poly(\log(N)]$ steps. Alternatively, we can resort to quantum random access memory\cite{2008-giovannetti-Qram,2008-giovannetti-ArchitectureQram,arunachalam2015robustness} or Hamiltonian simulation method\cite{2018-Wossning-QLSA}. Quantum Random Access Memory (qRAM) is an efficient method to do state preparation, whose complexity is $\mathcal{O}(\log(N))$ after the quantum memory cell was established.

As for the entire system, with the ancillary register $s$, $d$ being in a specific superposition state by $V_0$ and controlled-$V$, it is driven into
\begin{eqnarray}\label{Superstate}
|\psi_1\rangle= \frac{1} {\sqrt{\beta}} \left ( |0\rangle_s|0\rangle_d^{T_{1}} +\sum_{m=0}^{Kp-1} \sqrt{c_{m}}|1\rangle_s|m\rangle_d \right)|\vecx\rangle
\end{eqnarray}
with $\beta=1+\sum c_m$ and unitary matrixes $V$, $V_0$
\begin{eqnarray}
  V=\left(                 
  \begin{array}{cccc}   
    \sqrt{c_0} & v_{0,1}  & \cdots & v_{0,{Kp-1}} \\ 
         \vdots & \vdots  & \ddots  & \vdots \\ 
	             \sqrt{c_{Kp-1}} & v_{Kp-1,1} & \cdots & v_{{Kp-1},{Kp-1}} \\ 
        \end{array}
\right)
\end{eqnarray}
\begin{eqnarray}
 V_{0}=\left(                 
  \begin{array}{cc}   
    \frac{1}{\sqrt{\beta}} & \frac{\sqrt{\beta-1}}{\sqrt{\beta}} \\ 
        \frac{\sqrt{\beta-1}}{\sqrt{\beta}} &  -\frac{1}{\sqrt{\beta}}  \\ 
        \end{array}
\right).
  \end{eqnarray} 
Remarkably, $c_i$ should be rescaled with $c_i/\sqrt{\beta-1}$ to the unitary condition.  While all other elements $\{V_{0,1},V_{0,2},\cdots,V_{Kp-1,Kp-1}\}$  are arbitrary as long as $V$ is unitary.

Step 2: To apply the gradient operation, $\opD$, on the system, the methods of linear combination of the unitary operations are employed
\cite{2006-long-DQC,2006-Long-DQC2,childs2012, 2016-Wei-DQC1,2016-Wei-DQC2,wen2020one,xin2019preparation}.
$\matA_0$, $\matA_1$...$\matA_{Kp}$, tensor decompositions of $\matA$, are applied to the work system conditionally on the register $d$ which is on $|0\rangle$, $|1\rangle$...$|Kp-1\rangle$, correspondingly. In this way, the work system would feel an effective operation as $\sum_{i=0}^{Kp} \matA_{i}$ when registers $s$, $d$ are delicately decoupled. However, $\matA_0$ would be applied to the work system in both $|0\rangle_s|0\rangle_d$ and $|1\rangle_s|0\rangle_d$ subspaces. Thus, an additional $\matA_0^{\dagger}$ is required for the compensation and the final state is  
\begin{eqnarray}\label{Superstate}
|\psi_2\rangle= \frac{1} {\sqrt{\beta}} \left ( |0\rangle_s|0\rangle_d^{T_{1}}|\vecx\rangle +\sum_{m=0}^{Kp-1} \sqrt{c_{m}}|1\rangle_s|m\rangle_d \matA_m |\vecx\rangle  \right)
\end{eqnarray}

Step 3: Combination is implemented to combine the information in different subspaces of the ancillary system and generate the formalized $\opD$ on the work space. Controlled-$W$ and $W_0$, which are the inverse operations of $V$ and $V_0$, are applied in this step, which produces 

\begin{eqnarray}\label{psi3}
|\psi_3\rangle & = &\frac{1}{\beta}|0\rangle_s|0\rangle_d^{T_{1}}(|\vecx\rangle+\sum_{m=0}^{Kp-1} c_{m}\matA_m  |\vecx\rangle)\nonumber \\
 +&&|1\rangle_s|0\rangle_d^{T_{1}}(\frac{\sqrt{\beta-1}}{\beta}|\vecx\rangle- \frac{1}{\beta\sqrt{\beta-1}}\sum_{m=0}^{Kp-1} c_{m}\matA_m  |\vecx\rangle)\\
+&&(\frac{\sqrt{\beta-1}}{\beta}|0\rangle_s- \frac{1}{\beta}|1\rangle_s)\sum_{l=1}^{Kp-1}|l\rangle_d\sum_{m=0}^{Kp-1} v_{m,l}\sqrt{c_{m}}\matA_m  |\vecx\rangle \nonumber
\end{eqnarray}
The last two terms are orthogonal to the first term, and can be regarded as rubbish terms.
When the ancillary system is in state $|0\rangle_s|0\rangle_d^{T_{1}}$, the iterative state, $|\vecx'\rangle$(shown in Eq.(\ref{Exp_state})) is  obtained from the work system. 

Finally, the result state $|\vecx'\rangle$ could either be the answer state which satisfies the previous set-up convergence condition or be the next input for both parameter and iteration circuits. More details on the protocol can be found in supplementary note \textrm{I}(A) and (B)\cite{supple}

\emph{Algorithm complexity} is concerned with the computing resources required to process a quantum information task. Particularly, the gate complexity quantifies the amount of the basic quantum operations taken to run as a function of size of input. Similarly, the memory complexity quantifies the amount of space or memory taken.

$Kp$ is an important indicator to characterize the complexity of the protocol, where $K$ is the number of decomposition terms required to specify $\matA$ and $p$ is half the order of the cost function. It determines both the size of the ancillary system and the depth of both parameter and iteration circuits. 

For the size of the circuits which does not include state preparation, $\mathcal{O}(T_1+\log(N))$ qubits are required for both parameter and iteration circuits, where $T_1$ is the integer that satisfies $2^{T_1}= Kp$. And two copies of the iterative state are needed for each iteration. If the number of iterations is $r$, the total memory consumption is $\mathcal{O}(2^r \log(N)+2T_1)$.

For the depth of the circuits which already has the encoded states, $\mathcal{O}(Kp)$ conditional-$A_{m}$s are required. In addition, the gates complexity is provided under the assumption of Pauli product form of $A_{m}$. As a $\log(Kp)$-qubit-controlled gate can be implemented with $\mathcal{O}(\log(Kp)^2\times \log(N))$ basic quantum gates, which is included in supplementary note \textrm{I}(C) \cite{supple,2017-XIn-QSquantumchannel}, $\mathcal{O}(Kp\times \log(Kp)^2\times \log(N)\times r)$ basic quantum gates are required for both circuits for $r$ iterations.

As for the the state preparation step, the amplitude encoding method would consume $\mathcal{O}(\log(N))$ more qubit with $\mathcal{O}(poly\log(N))$ steps. 
If the qRAM is adopted in the state preparation, the spatial cost is not just $\mathcal{O}(\log(N))$ qubits, one also needs $\mathcal{O}(N)$ qutrits to establish quantum memory cell.

The protocol relies on the tensor decomposition of $A$, which is in general hard, especially as $Kp$ grows. This protocol is theoretically efficient when there is an explicit decomposition of $A$ with a limited $Kp$. However, there are some benefits when adopting this experimental protocol. The experiments are comparably easier since only two copies are required for each iteration optimization. 

\emph{Success probability} 
For the parameter circuit,  the probability of obtaining the required $b_j^{\alpha}$ is related to the size of the second ancillary resister, $T_1$, which is proportional to $1/Kp$.
For the iteration circuit,  the ancillary register finally stay on $|0\rangle_s|0\rangle^{T_{1}}_d$ and the output is determined to be the iterative state with the probability $P_{s}= \parallel|\vecx^{(t)}\rangle \pm D|\vecx^{(t)}\rangle\parallel^{2}/{(\sum_{m=0}^{Kp-1} c_{m}+1)^{2}}$.

\subsection*{Apparatus}
All experiments were carried out on a Bruker DRX 400MHz Nuclear Magnetic Resonance(NMR) spectrometer at room temperature. As it is shown in Fig.\ref{fig2}(a), a $4$-qubit system is required, represented by the liquid $^{13}$C-labeled crotonic acid sample dissolved in d-chloroform. Four carbon-13 nuclei spins($^{13}$C) are denoted as four qubits, $C_{1}$ as the register $s$, $C_{2,3}$ as the register $d$ and $C_4$ being the work system. The free evolution of this $4$-qubit system is dominated by the internal Hamiltonian, 

\begin{align}\label{Hamiltonian}
\mathcal{H}_{int}=\sum\limits_{j=1}^4 {\pi \nu _j } \sigma_z^j  + \sum\limits_{j < k,=1}^4 \frac{\pi}{2} J_{jk} \sigma_z^j \sigma_z^k,
\end{align}
where $\nu_j$ and $\emph{J}_{jk}$ are the resonance frequency of the \emph{j}th spin and the $J$-coupling strength between spins \emph{j} and \emph{k}, respectively. Values of all parameters can be found in the the experimental Hamiltonian of supplementary note \textrm{II}(A)\cite{supple}. 
In order to master the evolution of the system, the transverse radio-frequency(r.f) pulses are introduced as the control field,
\begin{eqnarray}
\label{RFHamiltonian}
\mathcal{H}_{rf}=-\frac{1}{2} \omega_1\sum_{i=1}^4 (\cos(\omega_{rf} t+\phi)\sigma_x^{i}+\sin(\omega_{rf} t+\phi)\sigma_y^i).
\end{eqnarray}
By tuning the parameters in r.f field(Eq.(\ref{RFHamiltonian})) such as intensity $\omega_1$, phase $\phi$ and frequency $\omega_{rf}$ and duration, the four-qubit universal quantum gates are theoretically achievable with the combination of internal system (Eq.(\ref{Hamiltonian}))\cite{2018-MQFC-Dawei,2019-Keren-PTensor}.

\subsection*{Experimental implementation}
A bivariate quartic polynomial(Eq.(\ref{objfun})), serving as the cost function, is shown to be minimized by our experimental protocol iteratively. The problem is depicted as
\begin{eqnarray}
 min \quad & f(\vecx)=\frac{1}{2}\vecx^{T}\otimes \vecx^{T}\matA \vecx\otimes \vecx 
\label{objfun}
\end{eqnarray}
with $|\vecx|=1$, where $\vecx=(x_1,x_2)^T$ is a 2-d real vector. Though the number of independent variable is $1$ for the normalization constrain, as the growth of the size of problem, a surge of information processing would be included. $\matA$, the coefficient matrix, has another representation by tensor products $\matA=-\sigma_I\otimes \sigma_x+\sigma_x\otimes \sigma_z$ , where $\sigma_{i} (i=I,x,y,z)$ denotes the Pauli matrices. 
 
With the amplitude encoding method($\vecx \rightarrow|\vecx\rangle$), the experimental demonstration for updating $|\vecx'\rangle$ consists of both acquiring parameters and proceeding iterations. The iteration circuit, from the $|0\rangle_s|00\rangle_d|\vecx\rangle$ to the output $|0\rangle_s|00\rangle_d|\vecx'\rangle$, is implemented with 3 steps, to $|\psi_1\rangle$, $|\psi_2\rangle$, $|\psi_3\rangle$ and measurement, sequentially. As for acquiring parameters, for the hermitian of $\matA_j^{\alpha}$, $c_m$ can be obtained as a measurement of $\matA_j^{\alpha}$ on $|\vecx\rangle \langle\vecx |$ instead of the parameter circuits. For accuracy and limitation of the molecule sample, this conversion is adopted and we concentrate on the iteration part, where $T_1$=$2$, $Kp$=$4$.

In the experiment, two sets of experiment $s_1$ and $s_2$ are conducted, with different initial guess, $\vecx^{s_1}_0$ $(-0.38, 0.92)$ and  $\vecx^{s_2}_0$ $(0.86, 0.50)$. The realization of the iteration circuit is depicted as follows:

\emph{Initialization} --- At room temperature,  the $4$-qubit quantum system is in the  thermal equilibrium state. This thermal equilibrium system can be driven to a pseudo-pure state(PPS) with spatial average method\cite{1997-Cory-NMRPPS}. Then, $|0\rangle_s|00\rangle_{d}|\vecx\rangle$ was prepared from this PPS, with simply a single-qubit rotation on $C_4$, where $\vecx$ is either initial guess $\vecx_0^{s_1}$, $\vecx_0^{s_2}$ or the output of last iteration. 
In this step, preparation of the PPS and individual control operations are the mature technology in NMR quantum control and can be found in supplementary note \textrm{II}(B)\cite{supple}.

\emph{Iteration circuit} --- The circuit consists of 3 steps and thus we pack our control pulses into 3 groups. They are shown in  Fig.\ref{fig2}(b): (1) A combination of single-qubit rotation $V_0$ and control-$V$ gate realizes the transformation to $|\psi_1\rangle$. (2) conditional operations of  decompositions of $A$ implement the $|\psi_2\rangle$. (3) $W_0$ and control-$W$ achieve the dis-entanglement to $|\psi_3\rangle$. Remarkably, parameters $c_{m}$ in local operations such as $V$ and $W$ are obtained by measuring the iterative state $|\vecx\rangle$. Gradient ascent pulse engineering (GRAPE) was employed to generate 3 packages of optimized pulses to implement the operations listed above, with the simulated fidelity all over 99.9\% and the time-length being 20ms, 30ms, 20ms, respectively \cite{2005-Khaneja-GRAPE}. Hence, in experiment, we got $\rho_1$,   $\rho_2$, and  $\rho_3$ correspondingly. 

\emph{Measurement} ---  Since only the state in subspace of $|0\rangle_s|00\rangle_{d}$ is necessary for obtaining the output, $|\vecx'\rangle$, a full tomography in such subspace was employed. All read out pulses are 0.9ms with 99.8\% simulated fidelity. For the sake of experimental errors, mixed states were led in our results, however,  the 2-dimension vector $|\vecx'\rangle$ should be a pure real state. Hence, a purification step was added to search a closest pure state after this measurement and it is realized with the method of maximum likelihood\cite{2001-Rehacek-reconEnt}. As the consequence of the output $|\vecx'\rangle$, $\ket{\phi_{i}^{s_1}}$ and $\ket{\phi_{i}^{s_2}}$ (i=1...4) were found to be the closest to our experimental density matrices for two different cases $s_1$ and $s_2$. 

According to pre-set threshold, the output $|\vecx'\rangle$ can be labeled as the updated input $|\vecx\rangle$ to run the next iterative circuit or  be the final result and the iteration thus terminates.

The results  are shown in Fig.\ref{fig3}. For the cases $s_1$ and $s_2$, we could see the trend of convergence at $\vecx_{opt}$$(0.50, 0.86)$ after 4 times iterations with the initial points, $\vecx^{s_1}_0$ and $\vecx^{s_2}_0$. $\ket{\phi_{i}^{s_1}}$ and $\ket{\phi_{i}^{s_2}}$(i=1...4) are outputs of iteration circuit at $i$-th iteration, which are plotted in the sub-figures (a) and (b). For comparison, theoretical simulation is provided, whose inputs were chosen as the output of the last experimental iteration. 
 
In addition, by substituting $x_1$=$\cos{(\theta)}$ and $x_2=\sin{(\theta)}$, the cost function is rewritten as $f(\vecx)=-2\sin ^{3} \theta \cos \theta$. Thus the problem is reduced to a one-dimensional unconstrained optimization problem, where the extreme points lie at $\theta$=$0$, $\pi/{3}$. Among them, $\theta$=$0$ is unstable while $\pi/{3}$ is the stable local minimum.  To show this results explicitly, both iteration outputs and the value of the cost function are shown in Fig.\ref{fig3}(c). In this situation, the initial guesses are $\cos(\theta)=-0.38$($s_1$ and colored red) and $\cos(\theta)=-0.86$ ($s_2$ and colored blue), respectively. As with the growth of the number of the iteration, the value of the cost function gets lower and lower, until slipping into the neighbor of the local minimum. 

As another aspect to show this convergence, in Fig.\ref{fig3}(d), relations between the number of iterations and overlaps were given. The value of vertical axis was defined as the overlaps between the optimal state and the output state after each iteration: $|\langle \phi_{opt} \ket{\phi_{i}^{\text{j}}}|$ (i=0,1..4. j=$s_1$,$s_2$). The horizontal axis is the number of iterations. It shows that the overlaps converges to 1  weather the initial guess is chosen as $\vecx^{s_1}_0$ or $\vecx^{s_2}_0$. For more information of the different seeds and investigation of unstable point, numerical simulations were carried out and some results are shown in supplementary note \textrm{III}\cite{supple}. 

Furthermore, to check the performance of the circuits experimentally implemented, a 4-qubit tomography was implemented at two points, after PPS preparation and after the iteration circuit. Thus 4-qubit states $\rho_{pps}$ and $\rho_3$ were obtained. For the PPS we got a fidelity about 99.01\%, and for those 4-qubit states $\rho_3$, they have an average of 94\% fidelity. Detailed information is shown in the experimental part, supplementary note \textrm{II}(C) and (D)\cite{supple}.   

\subsection*{Application}
For the further applications, multidimensional scaling (MDS) is a technique, providing a visual representation of the pattern of proximities in a dataset. It is a common method of statistical analysis in sociology, quantitative psychology, marketing and so on. We apply our method to quantize an algorithm for fitting the simplest of multidimensional scaling models in major applications in the method.

Given a matrix $\matA={\delta_{ij}}$, which is nonnegative symmetric with zero diagonal. A set of number $\delta_{ij}$ is the data collected in a classical multi-dimensional scaling problem and $\delta_{ij}$ is the dissimilarity between objects $i $ and $j$. Representing $n$ objects as $n$ points via ignoring the objects size,
the dissimilarity of objects  $i $ and $j$ is approximately equal to the distance between points $i $ and $j$ .  The goal is to find $ n$ points in $m$ dimensions, denoted by $\vecx_1, \vecx_2, \cdot , \vecx_n$ to form a configuration  with coordinates in an $n \times m$ matrix $\vecX$.

When $m=3$, it is reduced to a molecular conformation problem\cite{1993-Glunt-MolecularMDS}, which plays an important role in chemical and biological fields. Let $d_{ij}(\vecX)$denotes the Euclidean distances between the points $\vecx_i $ and $\vecx_j$ . It follows that
\begin{eqnarray}
d_{ij}^{2}(\vecX)=(\vecx_i-\vecx_j)^T(\vecx_i -\vecx_j)
\end{eqnarray} 
We minimize the loss function,  defined as 
\begin{eqnarray}
f(\vecX)=1/2\sum_i\sum_jw_{ij}(d_{ij}(\vecX)-\delta_{ij})^2
\end{eqnarray} 
where $W={w_{ij}}$ is a symmetric weight matrix that can be used to code various supplementary information. The purpose of this algorithm is to find the most suitable information visualization configuration.
Now we map it to a quantum version. First,  the loss function is rewritten as\cite{1988-leeuw-MDS}
\begin{eqnarray}
f(\vecX)=1/2\sum_i\sum_jw_{ij}\delta_{ij}^2-2g(\vecX)+h^{2}(\vecX)
\end{eqnarray} 
 where 
 \begin{eqnarray}
g(\vecX)=1/2\sum_i\sum_jw_{ij}\delta_{ij}d_{ij}(\vecX)
\end{eqnarray} 
and 
 \begin{eqnarray}
h^{2}(\vecX)=1/2\sum_i\sum_jw_{ij}d_{ij}^2(\vecX)
\end{eqnarray} 
Thus, we only need to minimize $f'(\vecX)=-2g(\vecX)+h^{2}(\vecX)$.
$g(\vecX)$ and $h^{2}(\vecX)$ can be further expressed as a trace of some matrixes muiltiproduction.
We have $g(\vecX)= \Tr (\vecX^TB(\vecX)\vecX)$ with $B(\vecX)=1/2\sum_i\sum_jw_{ij}A_{ij}k_{ij}(\vecX)$, where
 \begin{eqnarray}
 k_{ij}(\vecX)=1/d_{ij}(\vecX), d_{ij}(\vecX)\neq 0\\
  k_{ij}(\vecX)=0, otherwise.
\end{eqnarray} 
Similarly, $h(\vecX)^2= \Tr(\vecX^TC(\vecX)\vecX)$ with $C(\vecX)=1/2\sum_i\sum_jw_{ij}A_{ij}$.
Then, we have
 \begin{eqnarray}
 f'(\vecX)=\Tr(\vecX^TD(\vecX)\vecX)
\end{eqnarray} 
where $D(\vecX)=C(\vecX)-2B(\vecX)$.
It should be noticed that here $\vecX$ is a $n \times m$ matrix. In order to represent $\vecX$ as quantum states,   we map it to  a sum of $m$ column vectors $\vecX_v$of  $\vecX$.
Now, we can apply our quantum gradient algorithm to minimize the objective function

 \begin{eqnarray}
 f'(\vecX)=\sum_m\Tr(\vecX_v^TD(\vecX)\vecX_v)
\end{eqnarray} 
In this special case, the function order $p=1$ and $D(x)$ is a symmetric matrix which is likely to be decomposed efficiently. It  potentially yields an exponential speed up over the classical algorithm in multidimensional scaling problems.

This protocol also provides potential applications in quantum control technology. For example, the cost function could be reduced to a quadratic optimization problem in the form of $f(x)=\langle\vecx| \matA|\vecx\rangle$. If the coefficient matrix $\matA$ is restricted to a density matrix, the objective function represents the overlap between $\matA$ and $|\vecx\rangle\langle\vecx|$. Thus, we can product a state $|\vecx\rangle$ closely enough to a density matrix $\matA$ by finding the maximum of $f(x)$.  It can be used as  a quantum method to prepare the specific state.

\section{Discussion}

In this article, an experiment-friendly protocol is proposed to implement the gradient algorithm.  
The protocol provides a quantum circuit only comprised of standard quantum gates, hence it can in principle be realized in current technologies. The experimental implementation required  only two copies of quantum states for the parameter and iteration circuits. Moreover, if there is an explicit decomposition of $A$ in terms of Pauli product matrix, $\mathcal{O}(Kp)\times \log(N)$ depth circuits ($\mathcal{O}(Kp\times \log(Kp)^2\times \log(N))$ basic quantum gates) are enough to calculate the gradient within with $\mathcal{O}(T_1+\log(N))$ qubits. 

\begin{figure}[!ht]
  \centering
  \includegraphics[width= 0.94 \columnwidth]{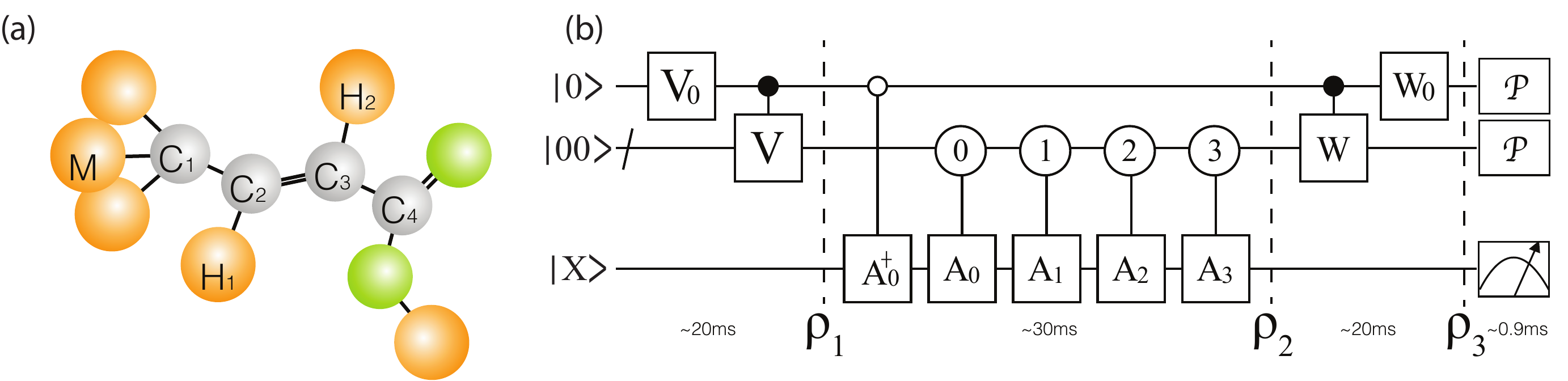}
  \caption{Molecule and quantum circuit. (a) Molecule structure of 4-qubit sample : crotonic acid. (b) Quantum circuit for an iteration to realize gradient descent algorithm.  $|\vecx\rangle$ denotes the initial state of work system, and  ancillary system are $T_{1}+1 $ qubits in the $|0\rangle|0\rangle^{T_{1}}$ state, where$ T_{1}=2$.}
  \label{fig2}
\end{figure}

\begin{figure}[!ht]
  \centering
  \includegraphics[width= 0.94\columnwidth]{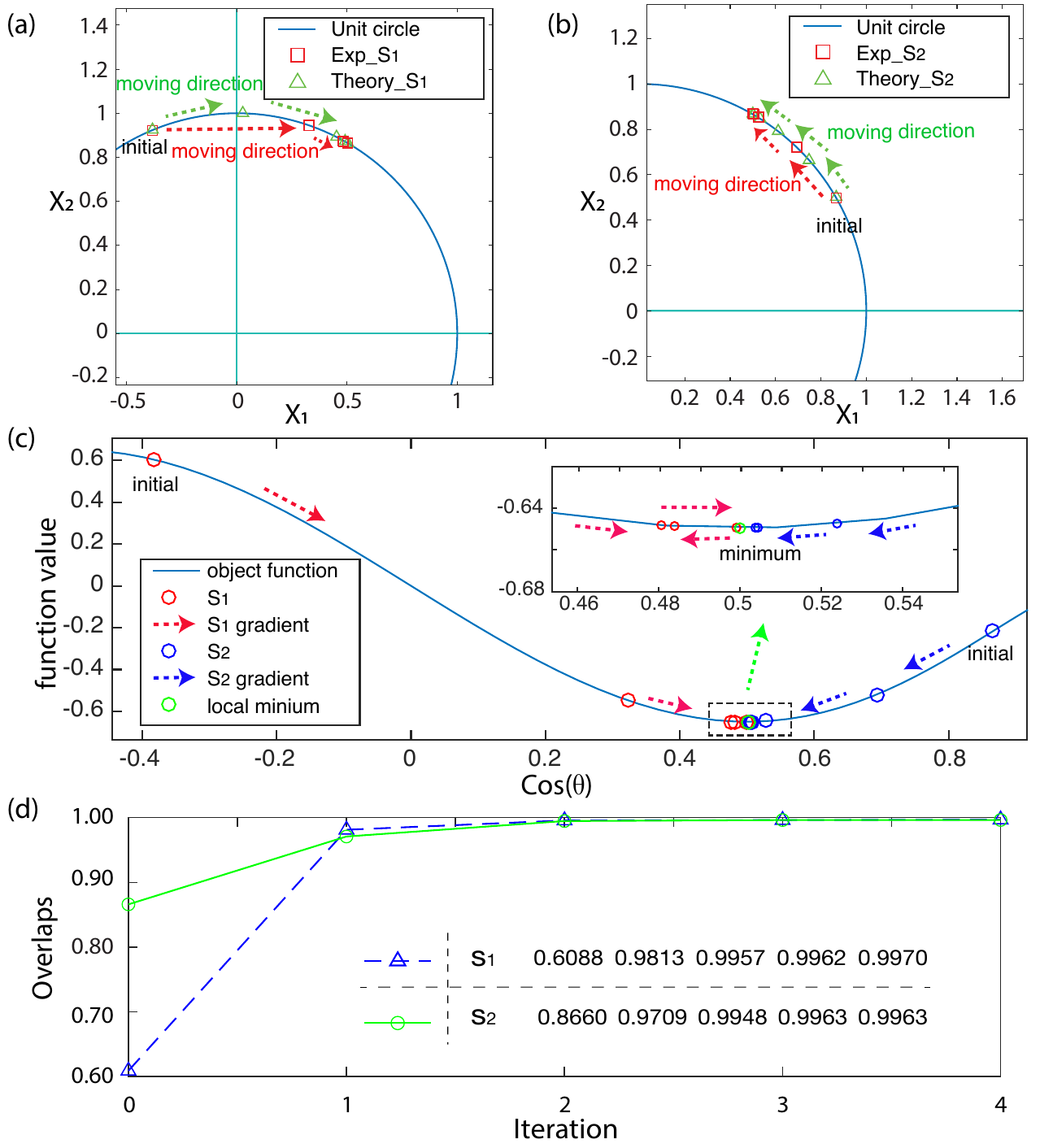}
  \caption{Theoretical simulation and experimental results: (a) and (b) show the output $|x'\rangle$ in the iteration process with their orthogonal basis. ie. $\ket{\phi_{i}^{s_1}}$ and $\ket{\phi_{i}^{s_2}}$(i=1...4). Green triangles are theoretical simulation results  while red squares are experimental measured outputs. They both begin with a same initial point. In addition, the moving directions are also labeled by the dashed arrows coloured green or red. (c) is the 1-d depiction. Beginning with two initial points, for $s_1$(coloured red) and $s_2$(coloured blue), the iteration outputs become lower and lower, until slipping into the neighbor of the local minimum. The dashed arrows show the moving direction for each iteration and in the zoom-in figure, it shows they gradually converging to the optimal minimum point were $x_1$ (or $\cos \theta$) = $0.5$.(d) gives the relations between the number of iterations and the overlaps between the iterative states and the target.}
  \label{fig3}
\end{figure}

With a 4-qubit NMR quantum system, we demonstrated an optimization of a homogeneous polynomial optimization and iteratively obtained the vicinity of the local minimum. The result is iteratively implemented with the iteration circuit in Fig.\ref{fig2} while the parameters $c_m$ are measured with iterative states instead of the parameter circuit. With the initial guess either $\vecx^{s_1}_0$ or $\vecx^{s_2}_0$, the demonstration shows the feasibility in near-future quantum devices for this shallow circuit. For the advanced control techniques of spin systems\cite{2005-ReviewNMR-Vandersypen,2019-Overcome-long}, they are applied as the first trail to demonstrate the effectiveness of the more and more protocols. In addition, Multi-Dimensional Scaling(MDS) problems are introduced as a potential application of this  experimental protocol.

Polynomials, subject to some constraints, are basic models in the area of optimization. Furthermore, the gradient algorithm is considered as one of the most fundamental solutions to those non-convex optimization problems. Our protocol, which gives another implementation of the gradient algorithm using quantum mechanics, is applicable to homogeneous polynomials optimization with spherical constrains. When there is a simple and explicit decomposition of coefficients matrix, the protocol could provide an speed-up with poly-logarithmic operations of the size of problem to calculate the gradient, which has potential to be used in near-future quantum machine learning. Our approach could be exceptionally useful for high-dimensional optimization problems, and the  gate-based circuit makes it readily transferable to other systems such as superconducting circuits and trapped ion quantum system, being an subroutine for future practical large-scale quantum computers.

\bigskip\noindent
\textbf{Data availability}
All data for the figure and table are available on request. All other data about experiments are available upon reasonable request.

\bigskip\noindent
\textbf{Code availability}
Code used for generate the quantum circuit and implementing the experiment is available on reasonable request.

\bigskip\noindent
\textbf{Acknowledgements}
This research was supported by National Basic Research Program of China.
K.L. acknowledge the National Natural Science Foundation of China under grant No. 11905111. 
S.W. F.Z. P.G. Z.Z. T.X. and G.L. acknowledge the National Natural Science Foundation of China under Grants No. 11974205, and No. 11774197. The National Key Research and  Development Program of China (2017YFA0303700); The Key Research and  Development Program of Guangdong province (2018B030325002); Beijing Advanced Innovation Center for Future Chip (ICFC). S.W. is also supported by the China Postdoctoral Science Foundation 2020M670172. T.X. is also supported by the National Natural Science Foundation of China (Grants No. 11905099 and No. U1801661), and Guangdong Basic and Applied Basic Research Foundation (Grant No. 2019A1515011383). X. W. was partially supported by the National Key Research and  Development Program of China, Grant No. 2018YFA0306703.

\bigskip\noindent
\textbf{Author Contributions}
G.L. initiated the project. S.W. formulated the theory. K.L. P.G. and Z.Z. performed the calculation. K.L. F.Z. and T.X. designed and carried out the experiments. All work was carried out under the supervision of  P.R. X.W. and G.L. All authors contributed to writing the manuscript. K.L. and S.W. contribute equally to this work.

\bigskip\noindent
\textbf{Competing interests}
The authors declare no competing interests.

\bibliographystyle{unsrt}

\newpage

\end{document}